\documentstyle[aps,preprint]{revtex}

\begin{document}

\draft

\title{Search for low--dimensional chaos in a reversed field pinch plasma}

\author{F. Sattin${}^{(1,2)}$\thanks{sattin@igi.pd.cnr.it} and
        E. Martines${}^{(1)}$\thanks{martines@igi.pd.cnr.it}  
        }

\address{ ${}^{(1)}$Consorzio RFX, Corso Stati Uniti 4, 35127 Padova, Italy}

\address{ ${}^{(2)}$Istituto Nazionale per la Fisica della Materia, 
           Corso Stati Uniti 4, 35127 Padova, Italy}

\maketitle

\abstract{An analysis of experimental data from the RFX (Reversed Field 
eXperiment) reversed field pinch machine [L. Fellin, 
P. Kusstatscher and G. Rostagni, Fusion Eng. Des. {\bf 25}, 315 (1995)]
is carried out to investigate the possible existence of 
deterministic chaos in the edge plasma region. The mathematical tools 
used include Lyapunov exponents, Kaplan--Yorke dimension, minimum
embedding dimension estimates and nonlinear forecasting. The whole analysis
agrees in ruling out the possibility of low--dimensional chaos: The 
dimension of the underlying dynamical system is estimated to be $ > 10$. From 
a critical re--reading of the literature it emerges that the  
findings of this work are likely to be common to all reversed field 
pinches.}


\section{Introduction}
Fusion machines are complex systems possessing a huge number of 
degrees of freedom where phenomena involve many scales of 
length and time. Such devices are therefore natural candidates for 
checking of nonlinear theories. In the past years, as soon as the 
experimental instrumentation made it possible,  
some effort has been exerted to determine if the temporal behaviour of 
measured quantities in fusion plasmas, displaying 
a fluctuating behaviour, 
could be modelled using some low--dimensional deterministic theory or 
were the manifestation of truly random processes 
\cite{persson,gee,cote,barkley,strohlein,watts1,watts2,sawley}. 
The results of this search for deterministic chaos are contrasted: 
while the results 
of numerical simulations \cite{persson} show the existence of low--dimensional 
chaos, experimental works find  positive \cite{gee,cote,barkley,strohlein} 
as well as negative results \cite{watts2,sawley}.  
    
In this paper we present a study on this subject done on  
experimental data taken from the RFX experiment: 
a large (R = 2m, a = 0.457m) toroidal device for the magnetic confinement of 
plasmas in reversed field pinch (RFP) configuration \cite{rfx} built in 
Padova (Italy) and designed to reach a plasma current of 2 MA. 
The data refer to measurements made in the outer region of the plasma, 
inside the last closed magnetic surface. 
A number of mathematical tools has been applied to these data to 
detect traces of deterministic chaos. Furthermore, a critical discussion of the 
existing literature on the subject is done, in the 
light of the results found, and some general conclusions are drawn.   
 
In the next section we illustrate the mathematical apparatus used, 
then a brief description of the diagnostics used to collect the data 
is given. Section IV presents the results of the analysis and finally 
some comments are given.  
  
\section{Numerical techniques}
Several analysis techniques have been developed to identify the presence 
of a low dimensional attractor in a dynamical system from its time series. 
In this section we describe the mathematical tools used in this work. Some of 
them are very recent and only recently have been applied to real experimental 
data. 
Therefore there is a further interest in verifying how they perform 
in this situation.  
All the techniques presented herein apply to a scalar time series of 
recorded  data $ S = (x_1 , x_2 , \dots , x_N)$. 
From the scalar series $S$ one may construct $q$--dimensional vectors 
in delay--coordinates,
\begin{equation}
{\bf X}_i(q) = [ x_i, x_{i+\tau} , \dots , x_{i+(q-1)\tau} ]	\quad .
\protect\label{eq:timedelay}
\end{equation}
The $q$--dimensional space of delay vectors plays a central role in 
this kind of analysis. The integer $\tau$ is known as the time lag. 
It can be larger than 1, {\it i.e.} data may be sampled at a frequency 
lower than the experimental one.
This is done since choosing $\tau$ smaller than the autocorrelation time 
$t_{c}$ of the data would introduce spurious correlations. On the other hand 
$\tau$ should neither be too large. The standard choice is to put
$ \tau \approx 2 \div 3 \,t_{c}$. In our units, depending upon the data,
$t_{c} = 1 \div 2$. To test the reliability of our results, $\tau$ 
was varied in the test from 1 to 7 times $t_{c}$. Conclusions were 
unaffected by the choice.

\subsection{Estimate of the minimum embedding dimension}
\label{sec:embedding}
In experiments, the true dimension of the phase space of the system under study 
(the embedding dimension $m$) is usually unknown. A reasonable guess for 
it is however essential for any analysis. It has been proven by Takens 
\cite{takens} that for a chaotic dynamical system the time recording 
of a single variable is sufficient to reconstruct the relevant dynamics--and 
in particular the dimension $d$ of the attractor--provided that $m > 2 
d + 1$ (in actual cases it is known that this condition may be slightly
relaxed). 

In this work we have implemented the method proposed by Cao \cite{cao} to 
estimate the minimum embedding dimension from a scalar time series. \\
This method uses the time--delay vectors (with $\tau$ set to one for 
simplicity)
\begin{equation}
{\bf y}_i(m) = [ x_i, x_{i + 1}, x_{i + 2 }, \dots , x_{i + (m-1)}]
\;, \;\; i = 1, 2, \dots , N - (m-1) 
\end{equation}
(where $m$ is the guessed embedding dimension) to build the function  
\begin{equation}
a(i,m) = { || {\bf y}_i(m+1) - {\bf y}_{n(i,m)}(m+1)|| \over  
	|| {\bf y}_i(m) - {\bf y}_{n(i,m)}(m)|| } \; , \;
	i = 1,2,\dots , N - m  \, ,
\end{equation}
where $||\cdot||$ is a norm and $n(i,m)$ is an integer such that 
$y_{n(i,m)}(m)$ is the nearest neighbour of ${\bf y}_i(m)$ in the 
$m$--dimensional space, according to the $||\cdot||$ norm. The actual functional 
form of $ ||\cdot|| $  does not appear to be of importance. We have used the 
maximum norm:
\begin{equation}
||{\bf Y} - {\bf Z} || = \max |Y_{i} - Z_{i}| \; , \; i = 1,\dots,m 
\quad .
\end{equation}
If $m$ is the true embedding dimension, any two points which are close 
together in the 
$m$--dimensional reconstructed space, will stay close also in the 
$m+1$ space. Points which satisfy this condition are called {\sl true neighbours}, 
otherwise {\sl false neighbours} \cite{kennel}. Starting from a low value 
for $m$ and approaching the correct value (hereafter referred to as $\nu$) the 
number of false neighbours should decrease to zero,  or equivalently $a(i,m)$ 
should reach a constant value. 
Cao \cite{cao} suggests to use averages of this quantity:
\begin{equation}
E(m) = { 1 \over N - m } \sum^{N - m }_{i = 1} a(i,m)
\end{equation}
and
\begin{equation}
E1(m) = E(m+1)/E(m) \quad ,
\protect\label{eq:euno}
\end{equation}
which allow to obtain results independent upon  the sample data chosen. 
$E1(m)$ should stop changing when $m$ becomes greater than some value $\nu$, 
if the time series has a finite dimensional attractor. 
In the case of random data $E1$ will never saturate but when dealing 
with real data it is difficult to distinguish if it has attained a constant 
value or is slowly increasing, therefore in Ref. \cite{cao} it is recommended 
to also compute the function
\begin{equation}
E2(m) = E^*(m+1)/E^*(m)	\quad ,
\protect\label{eq:edue}
\end{equation}
\begin{equation}
E^*(m) = { 1 \over N - m } \sum^{N - m }_{i = 1} | x_{i + m } -
	x_{n(i,m) + m }| \quad ,
\end{equation}
where the meaning of $n(i,m)$ is the same as above. For random data 
$E2$ will stay close to 1: the $x$'s are now independent random variables, 
and therefore their average distance will be the same regardless of the 
space dimension $m$. For deterministic data, there will be a certain 
correlation between them which makes $E2$ a function of $m$.
As an illustration of the method, we plot $E1, E2$ in Figure 1 
for two sets of data: one time series is computed using the Mackey--Glass equation
\begin{equation}
{ d x(t) \over d t} = - 0.1 x(t) + { 0.2 x(t- \Delta) \over 
	1 + x(t - \Delta)^{10} }
\protect\label{eq:mackey}	 
\end{equation}
with $\Delta = 30$, which is known to describe the dynamics of a chaotic system 
with an attractor dimension of about 3.6 \cite{casdagli}. The other time series 
is generated using a random number generator. In the random data series $E2$ is 
always very close to one and $E1$ slowly converges to the same value. 
The plot of the deterministic map shows instead that both $E1$ and $E2$ reach 
the same value at $D \approx 6$; furthermore the behavior of $E1$ is not 
that of an asymptotic convergence to 1, but is more akin to the reaching 
of a threshold value. 

\subsection{Lyapunov exponents}
\label{sec:lyapunov}
The Lyapunov exponents, measuring the average divergence or 
convergence of orbits in phase space, are among the most frequently used 
quantities to ascertain the presence of chaos. 
Given a map ${\bf x}(t) = {\bf f}_{{\bf x}_0}(t), \,{\bf f} : {\bf R}^m \to 
{\bf R}^m$, the Lyapunov exponents are defined as 
\begin{equation}
\lambda^{(k)} = \lim_{t \to \infty} { 1 \over t} \ln |{ \partial
{\bf f}_{{\bf x}_0}(t) \over \partial x^{(k)} } |=
\lim_{t \to \infty} {1 \over t} || J_t^{(k)}|| \; , \; k = 1,\dots, m \quad ,
\protect\label{eq:explyap}
\end{equation}
where $ || J_t^{(k)}|| $ is the {\sl k}th eigenvalue of the 
{\sl m}--dimensional Jacobian. 
A necessary condition for a system to be chaotic is to have at least 
one positive exponent. 

Several well established techniques exist to compute the Lyapunov exponents 
for a system whose dynamical evolution is analytically known. Extracting 
them from an experimentally determined data set is much more difficult due to 
the limited length of the sample and to the presence of noise. 
Existing algorithms usually fit the experimental points to an analytical map 
$ g : {\bf R}^m \to {\bf R}$ such that 
\begin{equation}
g([x_i, \dots, x_{i+m-1}]) = x_{i+m}
\end{equation} 
or, which is the same, to the map
\begin{equation}
\tilde{g}([x_i, \dots, x_{i+m-1}]) = [x_{i+1}, \dots , x_{i+m}] \quad, 
\quad \tilde{g}: {\bf R}^m \to {\bf R}^m \quad .
\end{equation}
Under quite general conditions, the largest $\nu$ Lyapunov exponents 
of ${\bf f}$ and $\tilde{g}$ are the same \cite{brown,gencay}. 
Therefore one estimates the $\lambda$'s using standard techniques 
on $g, \tilde{g}$. In this work 
we have used two codes for estimating Lyapunov exponents. The former is 
the code developed by Watts \cite{watts1,watts2}, based upon the method 
by Briggs \cite{brown,briggs}, and already used on data from a 
magnetically confined plasma. In this method the time series is 
embedded in a delay 
space of given dimension, a number of nearest neighbours is found, and their 
trajectory is fit to an analytical function (usually a polynomial: in 
our runs we have used a polynomial of order 2). Then the Jacobian 
(\ref{eq:explyap}) may be obtained by analytical differentiation. 
The second code has been developed by one of us \cite{sattin} based upon 
a method by Gencay and Davis Dechert \cite{gencay}. Here, a single 
global fit is done by using logistic maps, {\it i.e.} 
\begin{equation} 
g({\bf X}_i) = \sum_l^L v_l = 
\sum_l^L {\beta_l \over 1 + \exp(- b_l - {\bf w}_l \cdot {\bf X}_i )  } \quad ,
\protect\label{eq:logistic}
\end{equation}
where $L$ is the number of functions $v$ and each ${\bf w}$ is an 
$m$--dimensional array. The $b_{l}$'s, $\beta_{l}$'s and ${\bf w}_{l}$'s 
are fitting parameters.
Logistic maps have some interesting features: they may fit arbitrarily 
well any analytical function as well as its derivative. In Ref. \cite{gencay} 
it is 
stated that this choice of functions has some advantages over the 
local polynomial approach in terms of stability of results in presence of 
noise and of fewer needed data points. On the other hand it requires a 
nonlinear fitting which is computationally more demanding. 

\subsection{The correlation and  Kaplan--Yorke dimensions}
\label{sec:dimension}
The correlation dimension $D_c$ gives another estimate of the  embedding 
dimension of the system. $D_c$ is defined as 
\begin{equation}
C_d(r) = \lim_{N \to \infty} {1 \over N^2} \sum^N_{i \neq j = 1}
H(r - |{\bf X}_i - {\bf X}_j|) \quad ,
\protect\label{eq:correlazione}
\end{equation}
where $H$ is the Heaviside function and the ${\bf X}$'s are defined in 
Eq.(\ref{eq:timedelay}). For $m > \nu, \lim_{r \to 0} C(r) \approx r^{\nu}$.\\
The correlation dimension has been and currently is a favourite tool to 
diagnose the presence of chaos in fusion plasmas \cite{barkley,sawley}. 

Instead of $D_c$ we have computed, using 
Watts's code \cite{watts2}, the Kaplan--Yorke (or Lyapunov) dimension 
\begin{equation}
D_{KY} = j + { \sum_{i=1}^j \lambda_i \over - \lambda_{j + 1} }
\end{equation}
where $j$ is the largest integer such that $ \sum_{i=1}^j \lambda_i > 0$, 
with the $\lambda$'s ordered as $ \lambda_1 > \lambda_2 > \dots $. 
There exist some conjectures \cite{russell} according to which the 
uguagliance $D_{KY} \approx D_c$ holds \\
$D_{KY}$ is much easier to compute than $D_c$, since it is only 
necessary to know the Lyapunov exponents, previously computed. This 
strength is at the same time a weakness when there may exist 
uncertainties about the correct value of the exponents. This is our 
case, however our analysis does not rely just on this single parameter
and as we shall see, all results corroborate the indications from 
$D_{KY}$.

\subsection{Nonlinear forecasting}
\label{sec:forecasting}
Given a time series of finite length representative of a chaotic 
dynamical system governed 
by the map $ {\bf f}: {\bf R}^m \to {\bf R}^m$, an inverse 
problem consists in finding a smooth map $ \tilde{g} : {\bf R}^m \to {\bf 
R}^m$, or its projection $ g : {\bf R}^m \to {\bf R}$ such that $\tilde{g}$ be 
an accurate approximation of ${\bf f}$. In the case $\tilde{g}$ is obtained, 
it may be used to predict accurately further data points. Otherwise, 
if the system is not governed by a finite dimensional map, or if the guessed 
$m$ is too low, the forecasting will be unreliable after few predictions. 
A review about the subject may be found in Ref. \cite{casdagli}. \\
We have used as fitting function the logistic maps (Eq. 
\ref{eq:logistic}): the original time series $S$ of length $N$ has been 
divided into two parts of lengths $N_1$ and $N_2 = N - N_{1}$. The first 
$N_1$ data have been used to fit Eq. (\ref{eq:logistic}) and the predictive 
power has been tested on the remaining $N_2$ points. The measure of goodness 
is the {\sl predictive error}: 
\begin{equation}
\sigma^2 = {1 \over N_2} \sum_{n=N_1+1}^N {\left( x_n - g({\bf X}_{n - m})
\right)^2 \over  \sigma^2_x } 
\protect\label{eq:sigma}
\end{equation}
where $ \sigma^2_x$ is the variance of the time series (in this work 
all data have been normalized so to have $ \sigma^2_x = 1$). The increase 
of the trial embedding dimension--provided that enough fitting parameters 
are allowed--will give a slight decrease of $\sigma^2$ until $ m \approx 
\nu$, when a sudden decrease to much smaller values is expected. 
To provide the reader with an example, in Figure 2 
we have plotted $\sigma^2$ {\it versus} $m$ for the Mackey--Glass map 
where is clearly visible the decrease of more than two orders of 
magnitude of $\sigma$ when the dimension is $ > 4$.  For comparison
we have estimated the predictive power of the method against a time 
series generated from a normal distribution. As expected the 
possibility of any forecast is null, with $\sigma^{2}$ always  greater 
than the variance of the original data.

It is worth mentioning that estimating the Lyapunov exponents and 
correlation dimensions has been attempted in all of this kind of studies 
concerning fusion plasmas. The 
forecasting of the data, however, has been applied, to our knowledge, only 
to data from the  Madison Simmetric Torus (MST) \cite{mst} reversed field 
pinch device \cite{watts2}. The estimate from the time series 
itself of the embedding dimension, finally, is applied in this work for the 
first time. 

Some words must be spent about the confidence which may be assigned 
to the algorithms. Two crucial topics affecting their performances
are: the number of data available and their quality, {\it i.e.} how 
much they are polluted with noise.
As far as the first point is concerned, usually the more data one 
can elaborate the larger may be the dimension of the system which may 
be correctly estimated. We had available records of
some thousands values, about as many as used by MST group. 
The authors of all algorithms used here claim them to be able 
to reconstruct  the correct 
dynamics of a low--dimensional system ({\i.e.} four or 
five--dimensional) using few hundreds data, with the 
possible exception of the Watts' code (see \cite{watts1,watts2}). 
In conclusion, and by comparison with the MST group's estimates, we may 
assert that dimensions up to or just below 10 may be correctly detected by our 
techniques.

\section{RFX experimental data} 
Signals coming from three different diagnostic techniques have been 
analysed in search of low dimensional chaos features: 

(a)  Floating potential ($V_f$): This is the potential of an electrically 
insulated conducting  probe immersed in the edge plasma. It is known to be 
related to the local plasma potential $V_p$ and to the local electron 
temperature $T_e$ (in eV) through the relationship $ V_f = V_p-\alpha T_e$ 
where  $\alpha$ is a constant which for the RFX edge 
plasma is approximately equal to 2.5. The probe was a graphite pin 
housed in a boron nitride structure which protected it from the 
unidirectional superthermal electron flow commonly observed in RFP edge 
plasmas. Data were sampled at 1 MHz. The measurements were collected 
during the experimental campaign described in \cite{prl}. 

(b) Time derivative of the radial magnetic field ($dB_r/dt$): It was  
measured with a pick-up coil housed in the same boron nitride 
structure as the $V_f$ measuring pin. This is a local measurement, like the 
previous one. Data were sampled at 1 MHz. 

(c) Density fluctuations at the edge: Collected by a reflectometer. 
The RFX reflectometer has been especially designed to  deal with 
high--frequency fluctuations. It is an homodyne reflectometer 
\cite{cavazzana} operating at a maximum sweep rate of 4 GHz/$\mu$s in 
the range 34-38 GHz. The collected signal is of the form 
\begin{equation}
s(t) = A(t) \cos(\Delta \phi (t) )
\end{equation}
with $A$ amplitude and $ \Delta \phi$ phase difference of the 
reflected radiation at a fixed microwave frequency. 
In this work we have analyzed the temporal behaviour of the amplitude. 

Figure 3 displays the plasma current waveform and typical waveforms for 
the three signals considered. The data were collected in discharges 
having a plasma current ranging between 350 and 600 kA.

\section{Results}

Figure 4 displays the quantities $E1, E2$ 
(Eqns. \ref{eq:euno}, \ref{eq:edue}) for the three analyzed signals. 
The data are taken for three different shots (shot 8422 for magnetic 
fluctuations, shot 7999 for potential measures, and shot 7852 for 
reflectometer data). A number of data points ranging up to about 10000 
has been used.  
The resemblance with random data (Fig. 1)  is impressive, 
which clearly suggests the existence of a very high 
dimensional phase space. 

A further confirmation is obtained by plotting the Kaplan--Yorke 
dimension (Fig. 5): no sign of saturation is obtained 
and over the whole explored range $D_{KY}(m) \approx m$. 

In Figure 6 we plot the predictive error 
(Eq. \ref{eq:sigma}). The value of $\sigma$ is always close to one, 
meaning that no real accurate prediction may be done. 
Some very small differences may be seen among the three signals, even 
if they may well be just subjective impressions: the magnetic 
fluctuations data show the worst predictive power, suggesting perhaps that 
some different mechanism from the other two diagnostics is at work. This 
could be the case if magnetic fluctuations are mostly due to 
magnetohydrodynamics tearing 
modes whereas potential and density fluctuations are mainly caused by some 
electrostatic instabilities localized in the edge region. 
For the numerical aspect, the calculation of this quantity turned out to be 
rather reliable, with limited variations between runs. 

Finally, in Figure 7 we plot the largest Lyapunov exponents. 
Since the system does not appear to be dominated by low--dimensional 
chaos, the usefulness of these coefficients is rather limited, 
however it may be interesting to compare the two approaches. 
The estimation of the $\lambda$'s from real data is known to be 
a difficult task; we did not obtain stable results varying the parameters: 
indeterminacies of $ 50 \%$ are quite likely, so the values shown are 
to be considered as representative of the general trend. We found 
that Watts's code is rather sensitive to the choice of the input 
parameters such as the order of the polynomial and the number of 
neighbours 
to be used in the fitting. Our code is more stable from this point 
of view since it needs very few input data (essentially, the embedding 
dimension $m$ and the number of logistic map $L$). However, it is 
difficult to perform the fit over a large number of data points (600 is 
the maximum used), so the results suffer of the scarce statistics. 

In Watts's work it is emphasized that, in order for the previous 
analysis to hold,  the physical system must be in a stationary state. 
The breaking down of this 
assumption may translate to an overestimate of the attractor dimension 
or even make it non measurable. 
Watts quotes two possible causes for the lack of non--stationarity: 
(a) The plasma may not 
reach a state of equilibrium;  (b) Even if equilibrium is reached, 
random perturbations (Watts cites as an example the influx of impurity 
ions from the walls) may destroy it. 
In RFX 
discharges true flat--top periods lasting some tens of milliseconds are reached.
In our study we have considered both discharges where this flat--top 
period was reached, and others where instead plasma parameters were 
slowly changing. Conclusions are unaffected by the discharge chosen.
We are therefore confident that any lack of stationarity due to this causes may
just very slightly modify the results.
Point (b) is by nature 
uncontrollable: however, since the same results are obtained from different 
discharges it may be hoped that these random events have not affected the 
final results. 

All our results agree in clearly pointing out that no low--dimensional 
chaos appears at the RFX edge. This is well consistent with findings of 
Watts {\it et al.} \cite{watts1,watts2} for MST. 
When limiting to RFP's the only 
other research within this field is that done in HBTX1A \cite{hbtx1a} by 
Gee and Taylor \cite{gee}, where 
traces of low--dimensional chaos were found in magnetic--field 
oscillations. 
This sharp discrepancy puzzled us, therefore  we resorted to check 
conclusions of Gee and Taylor against their own data. In Ref. \cite{gee} the 
attractor dimension is estimated 
using the correlation dimension technique (Eq. \ref{eq:correlazione}); 
their figure 1 shows the behaviour of $ \log C(r)$ {\it versus} $\log r$. 
Gee and Taylor claim that the slope of these lines saturate in correspondence 
of a dimension $ \approx 7$. We could not check this since in their figure 
1 only dimensions from 1 to 6 are plotted. However, from the data available, 
we could not find any clue of such a saturation: the slope of the straight 
line for $ m = 6$, as estimated by visual inspection of the plot, 
appears close to 8. Even allowing for the large error induced by our 
gross way of estimating, the true result cannot be much smaller 
than 6, which is a necessary condition to speak about a saturation of the 
slope \cite{avviso}. 

Up to this point, our work has focussed almost entirely on RFP plasmas. 
Some interesting considerations may be drawn by comparison with 
those tokamak plasmas where the fingerprints of deterministic 
chaos have been found. We refer to the works \cite{cote,barkley}. 
From Ref. \cite{cote}, using data of TOSCA \cite{tosca} and the 
Joint European Torus (JET) \cite{jet} tokamaks, it appears 
that a small value of the dimension of the dynamical system is 
more likely to be found if the level of turbulence (measured for example 
by $\delta B/B$) is small, which is not what happens in RFP's. Barkley's data 
\cite{barkley} 
are a bit difficult to interpret since their analysis are done on 
filtered data, {\it i.e.} by selecting the wave number components $k$.  
Their finding is that 
the dimension increases with  $k$. 
However, Barkley's data are chord averaged, so the central plasma plays a
dominant role. In our work, and in the others studied, only
edge quantities have been considered. It is quite possible that 
different mechanisms be at work in the two zones.

\section{Summary and conclusions}
In this work four statistical tools have been applied to the signals 
of some plasma turbulence measurements (magnetic field--, electrostatic 
potential--, and density fluctuations) of the RFX experiment, addressing the 
question of the existence of low dimensional deterministic chaos in them. 
The methods of analysis are well established (correlation dimension) 
as well as more recent (predicting errors and minimum embedding dimension 
estimates). \\ 
All the conclusions are strikingly in accordance in ruling out that the 
dynamics of the edge plasma in RFP's may be significantly affected by any 
low--dimensional process \cite{nota}. An estimate of a lower bound for 
the dimension of the system may be given by Figure 6, which shows that this 
value must be greater than 10. 
This conclusion is further strenghtened by a critical re--examination of 
previous results \cite{gee} conflicting  with ours, which has shed some 
doubt about their validity. 

\section*{Acknowledgements}
The authors are very grateful to Dr. C.A. Watts for allowing them to 
use his code, and to Dr. R. Cavazzana for providing the reflectometry 
data.

\begin{figure}
\label{embedding1}
\caption{Estimate of the minimum embedding dimension using Cao's method
(section \ref{sec:embedding}). }
\end{figure}

\begin{figure}
\label{fig:glassforecast}
\caption{Diamonds, predictive error for the Mackey--Glass equation (Eq. 
\ref{eq:mackey}) {\it versus } embedding dimension as estimated by 
the method  of section \ref{sec:forecasting}; Squares, the same for a 
series randomly generated from a normal distribution.}
\end{figure}

\begin{figure}
\label{fig:plasma}
\caption{Examples of the signals used. Smooth curves are the plasma 
current ($I_p$), the wildly fluctuating ones are the signals. Note that we
are referring to three different discharges.}
\end{figure}

\begin{figure}
\label{fig:expembedding}
\caption{$E1$, $E2$ (see section \ref{sec:embedding}) for the three signals. 
Here and in all the following plots $V$ stands for potential fluctuations; 
$B$, magnetic field; $n_e$, density fluctuations. Diamonds, $E1$; 
Stars, $E2$.}
\end{figure}

\begin{figure}
\label{fig:kydim}
\caption{Kaplan--Yorke dimension for the three signals.}
\end{figure}

\begin{figure}
\label{fig:esponenti}
\caption{ The largest  Lyapunov exponent {\it versus} embedding dimension.
Stars are results from Watts' code, diamonds from our code. }
\end{figure}

\begin{figure}
\label{fig:previsioni}
\caption{Predictive error {\it versus} embedding dimension.}
\end{figure}


\begin{references}

\bibitem{persson} M. Persson and H. Nordman, Phys. Rev. Lett. {\bf 
         67}, 3396 (1991).

\bibitem{gee} S.J. Gee and J.B. Taylor, in {\sl Proceedings of the 12th
	European Conference on Controlled Fusion and Plasma Physics},
        vol. II (European Physical Society, Geneva, 1985), p. 446.	

\bibitem{cote} A. Cot\'e, P. Haynes, A. Howling, A.W. Morris, and 
        D.C. Robinson, in {\sl Proceedings of the 12th
	European Conference on Controlled Fusion and Plasma Physics},
        vol. II (European Physical Society, Geneva, 1985), p. 450.
        
\bibitem{barkley} H.J. Barkley, J. Andreoletti, F. Gervais, J. 
         Olivain, A. Quemenur, and A. Truc, Plasma Phys. Contr. Fus.
	    {\bf 30}, 217 (1988). 

\bibitem{strohlein} G. Strohlein and A. Piel, Phys. Fluids B {\bf 1}, 
         1168 (1989).
         
\bibitem{watts1} C.A. Watts, {\sl Chaos and Simple Determinism in Reversed
	Field Pinch Plasmas}, Ph.D. thesis, University of Wisconsin (1993).

\bibitem{watts2} C.A. Watts, D.E. Newman, and J.C. Sprott, Phys. Rev. E
	{\bf 49}, 2291 (1994). 

\bibitem{sawley} M.L. Sawley, W. Simm, and A. Pochelon, Phys. Fluids {\bf 30}, 
	129 (1987).

\bibitem{rfx} L. Fellin, P. Kusstatscher and G. Rostagni, Fusion Eng. 
         Des. {\bf 25}, 315 (1995).

\bibitem{takens} F. Takens, in {\sl Dynamical Systems of Turbulence},
	edited by D.A. Rand and L.S. Young, Lecture Notes in Mathematics,
	Vol. 898 (Springer, Berlin, 1981), p. 366.

\bibitem{cao} L. Cao, Physica D {\bf 110}, 43 (1997).

\bibitem{kennel} M. Kennel, R. Brown, and H. Abarbanel, Phys. Rev. A {\bf 45},
	3403 (1992).
	
\bibitem{casdagli} M. Casdagli, Physica D {\bf 35}, 335 (1989).

\bibitem{brown} R. Brown, P. Bryant, H. Abarbanel, Phys. Rev. A {\bf 43},
	2787 (1991).

\bibitem{gencay} R. Gencay and W. Davis Dechert, Physica D {\bf 59}, 142 (1992).

\bibitem{briggs} K. Briggs, Phys. Lett. A {\bf 151}, 27 (1990).

\bibitem{sattin} F. Sattin, Comput. Phys. Commun. {\bf 107}, 253 (1997).

\bibitem{russell} D.A. Russell, J.D. Hansen, and E. Ott, Phys. Rev. Lett.
	{\bf 45}, 1175 (1980).

\bibitem{mst} R.N. Dexter, D.W. Kerst, T.W. Lovell, S.C. Prager, and J.C. 
Sprott, Fus. Techn. {\bf 19}, 131 (1991).

\bibitem{prl} V. Antoni, R. Cavazzana, D. Desideri, E. Martines, G. 
Serianni, L. Tramontin, Phys. Rev. Lett. {\bf 80}, 4185 (1998). 

\bibitem{cavazzana} R. Cavazzana, F. Chino, M. Moresco, A. Sardella, 
        and E. Spada,  in {\sl Proceedings of the 24th
	    European Conference on Controlled Fusion and Plasma Physics},
        vol. I (European Physical Society, Geneva, 1997), p. 361.

\bibitem{hbtx1a} H.A.B. Bodin, C.A. Bunting, P.G. Carolan, L. 
Giudicotti, C.W. Gowers, Y. Hirano, I.H. Hutchinson, P.A. Jones, C. 
Lamb, M. Malacarne, A.A. Newton, V.A. Piotrowicz, T. Shimada, M.R.C. 
Watts, in {\sl Proceedings of the 9th Conference on Plasma Physics 
and Controlled Nuclear Fusion Research}, vol. 1 (International Atomic 
Energy Agency, 1983), p. 641. 

\bibitem{avviso} Notice that also Watts (see Ref. \cite{watts1}) did 
	a critical re--reading of most of the papers dealing with
	chaos in fusion devices, reaching the conclusion that the 
	results shown in many of them suffered of an unappropriate
	elaboration of the data. Gee and Taylor's results, however,
	were not explicitly questioned by Watts.

\bibitem{tosca} K. McGuire, D.C. Robinson, A.J. Wootton, in
{\sl Proceedings of the 7th Conference on Plasma Physics and 
Controlled Nuclear Fusion}, vol. 1 (International Atomic 
Energy Agency, 1979), p. 335. 

\bibitem{jet} E. Bertolini, Fusion Eng. Des. {\bf 30}, 53 (1995).     

\bibitem{nota} This is not exactly the same as asserting that
{\sl no} low--dimensional processes exist. Actually, the signals may 
be composed by a contribution coming from these processes together 
with another due to turbulence. What we can say is that 
the latter is overwhelming.   
	
\end{references}
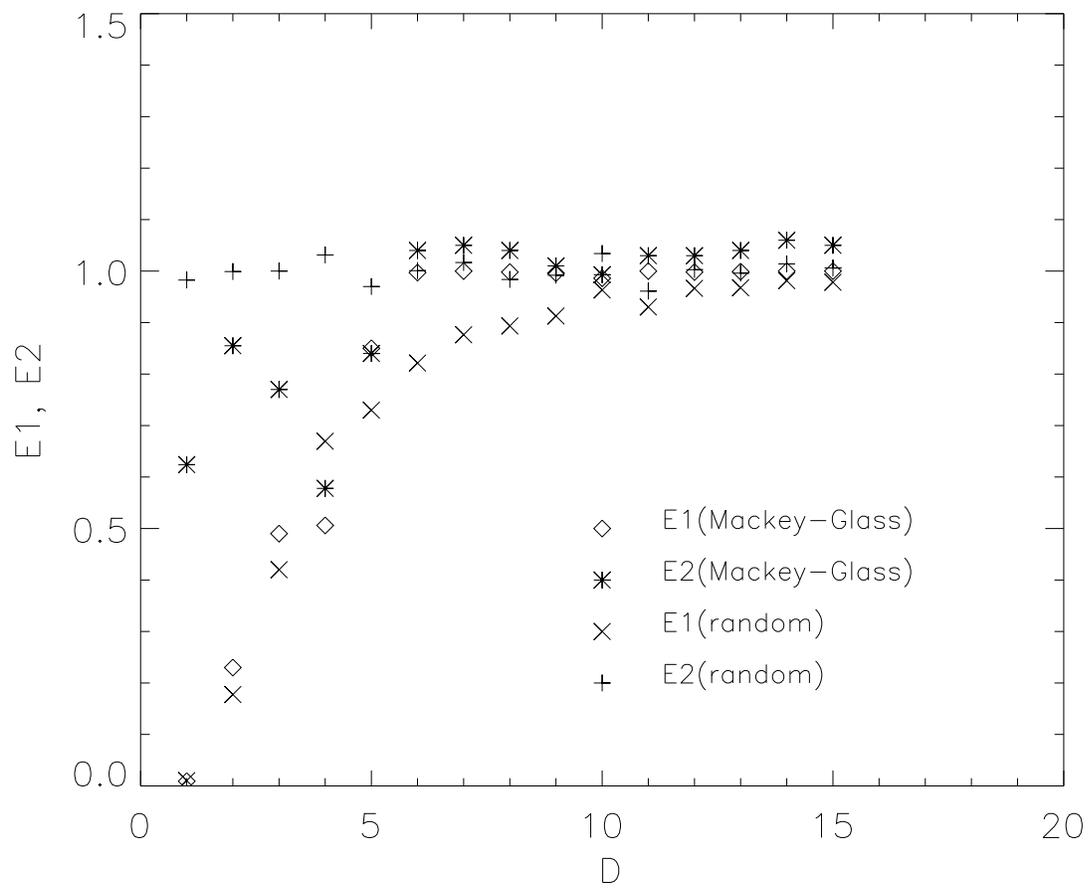
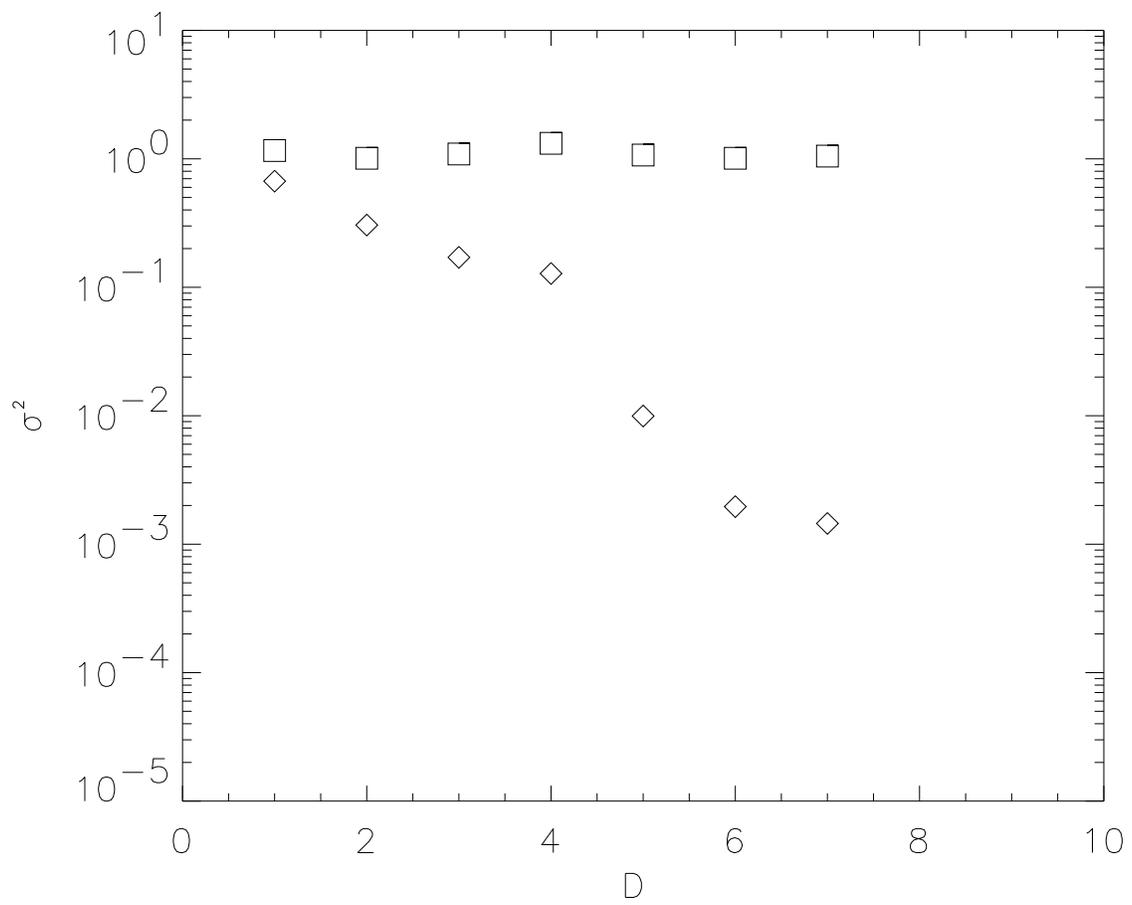
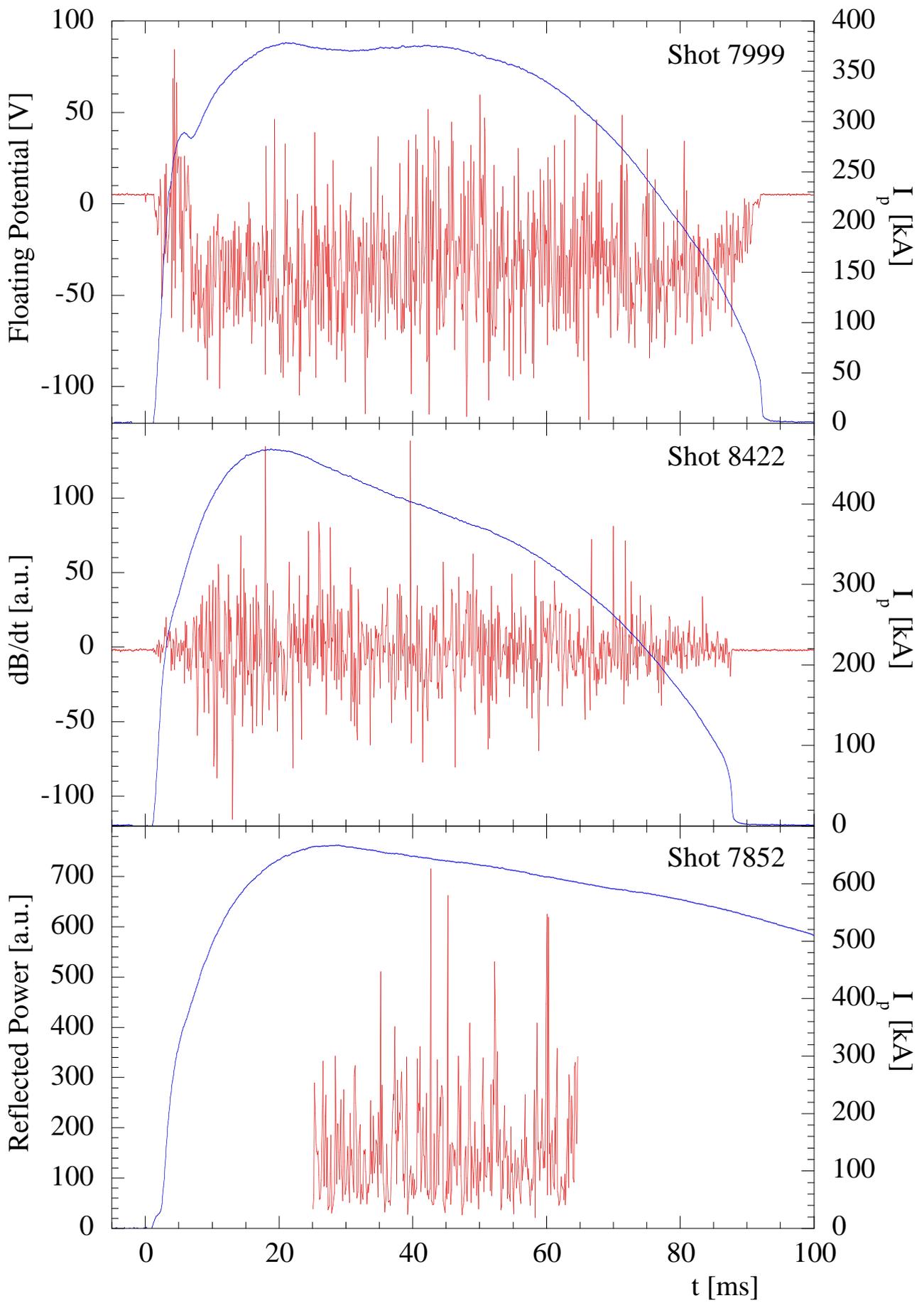
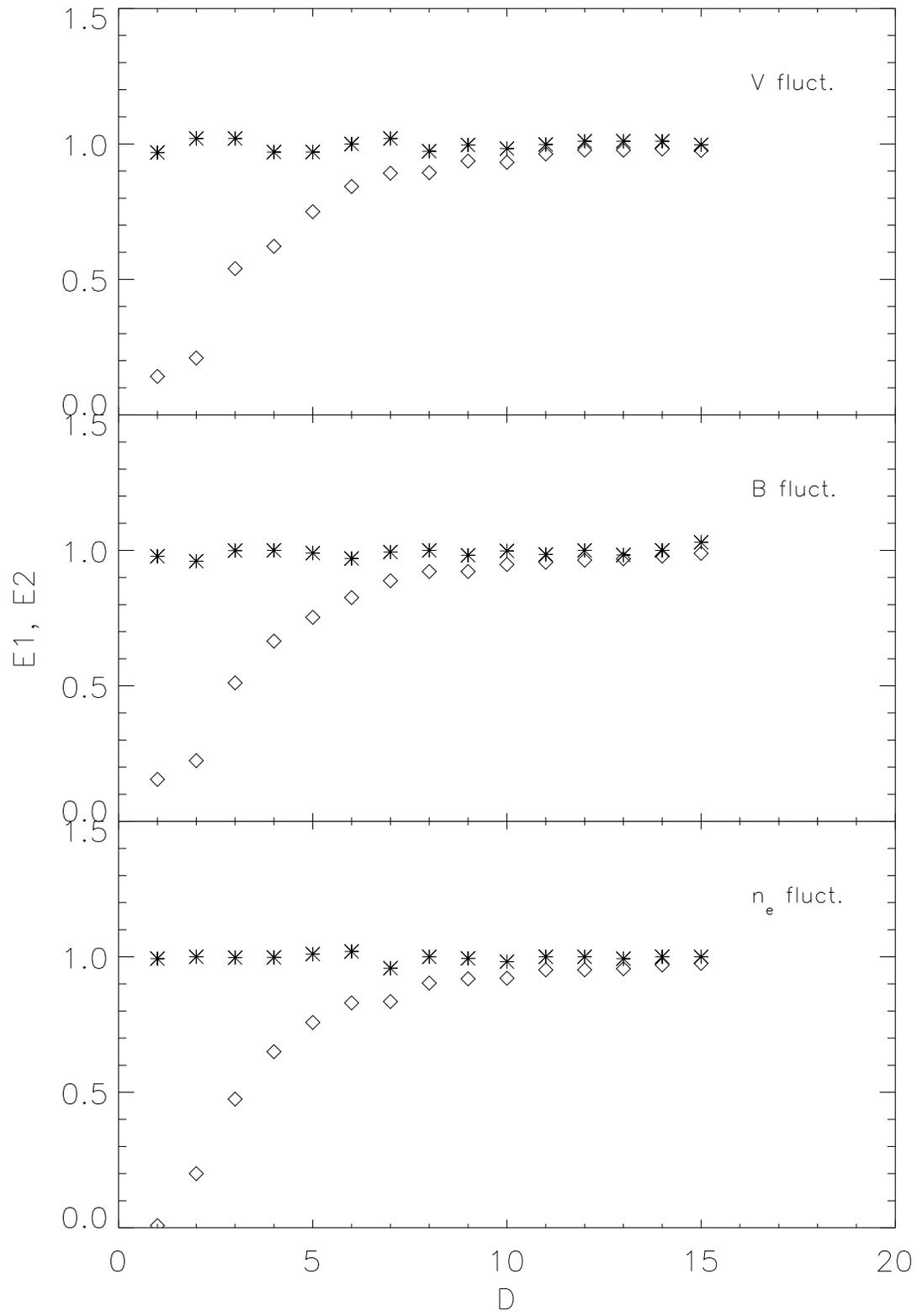
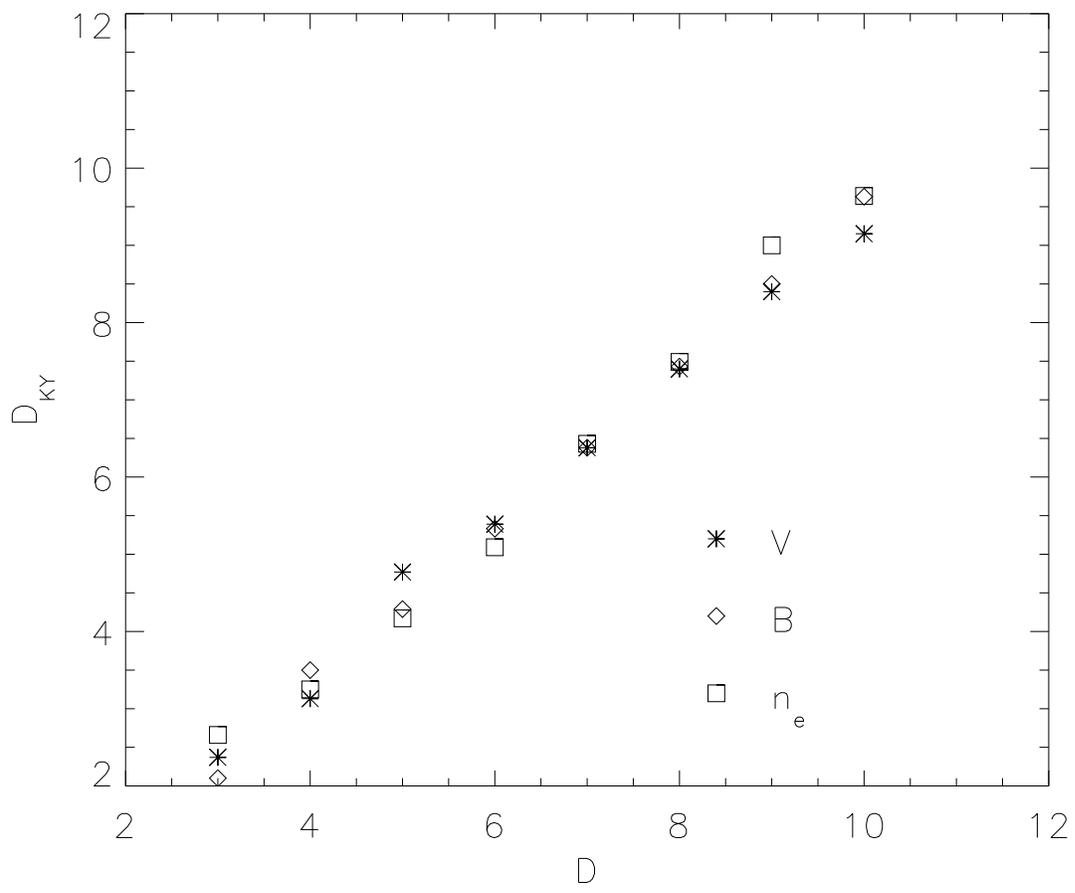
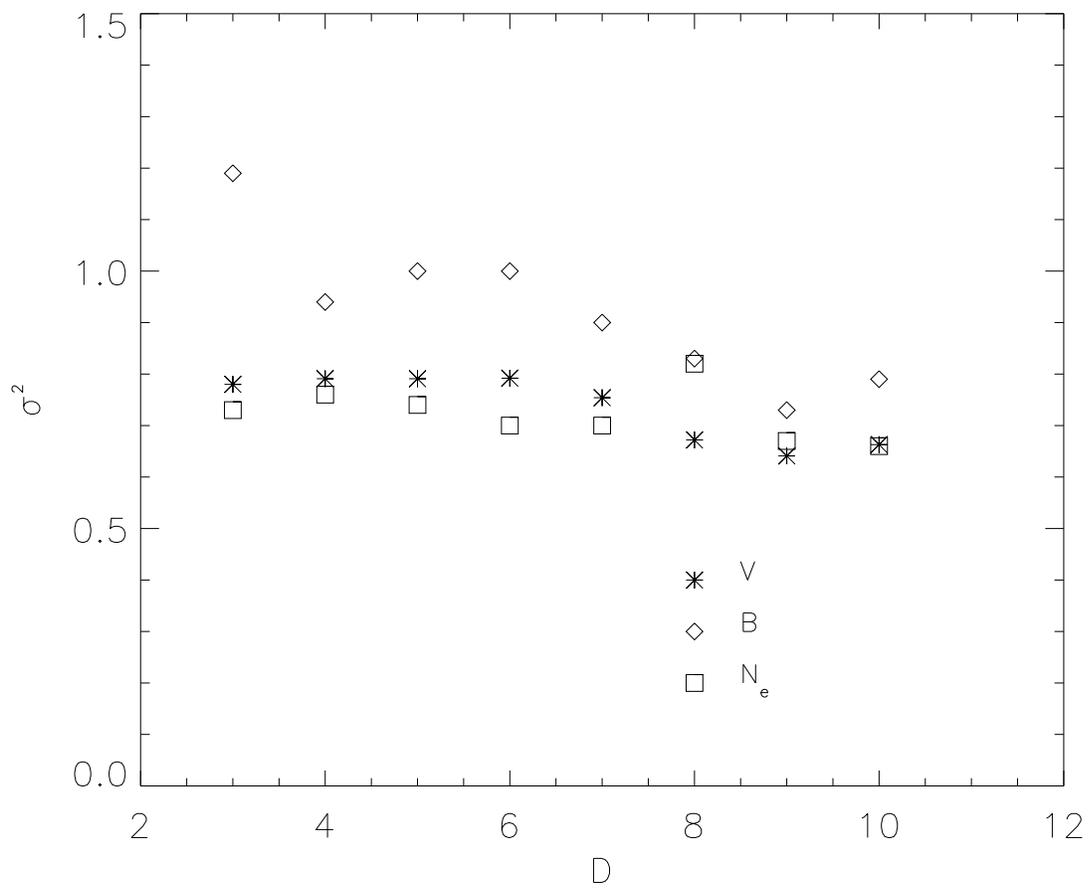
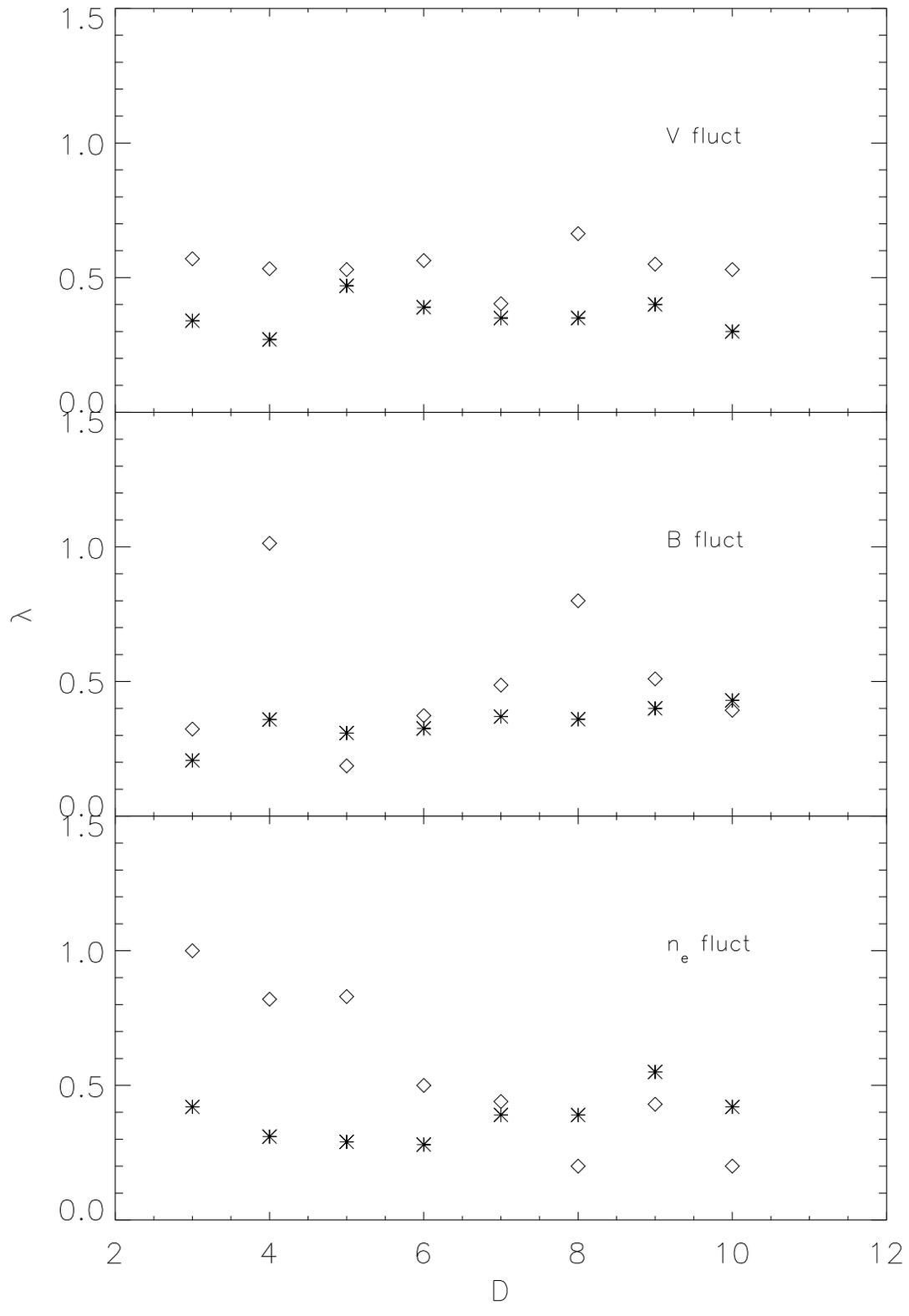
\end{document}